\documentclass[aps,prl,10pt,twocolumn,showpacs,superscriptaddress]{revtex4-1}
\usepackage{amsmath}
\usepackage{latexsym}
\usepackage{amssymb}
\usepackage{bm}
\usepackage{graphics,epstopdf}
\usepackage{color}
\usepackage{hyperref}

\usepackage{newlfont}
\usepackage{amsfonts}
\usepackage{amsthm}
\usepackage{graphicx}
\usepackage{epsfig}

\newcommand{\ket}[1]{|{#1}\rangle}
\newcommand{\bra}[1]{\langle{#1}|}
\newcommand{\braket}[2]{\langle {#1} | {#2} \rangle}

\usepackage{times}

\usepackage[up]{subfigure}

\newcommand{\be}{\begin{equation}}
\newcommand{\ee}{\end{equation}}
\newcommand{\bc}{\begin{center}}
\newcommand{\ec}{\end{center}}
\newcommand{\bea}{\begin{eqnarray}}
\newcommand{\eea}{\end{eqnarray}}
\newcommand{\ba}{\begin{array}}
\newcommand{\ea}{\end{array}}

\begin{document}
\title{Uncertainty and Complementarity Relations in Weak Measurement}

\author{Arun Kumar Pati}
\email{akpati@hri.res.in}

\affiliation{Department of Mathematics\\
Zhejiang University, Hangzhou 310027, PR~ China}

\affiliation{Quantum Information and Computation Group,\\
Harish-Chandra Research Institute, Chhatnag Road, Jhunsi,
Allahabad 211 019, India}

%\author{Lorenzo Maccone}
%\affiliation{ Dip.~Fisica and INFN Sez.~Pavia, \\
%University ~of Pavia, via Bassi 6, I-27100 Pavia, Italy }

%\author{Uttam Singh}
%\email{uttamsingh@hri.res.in}
%\affiliation{Quantum Information and Computation Group,\\
%Harish-Chandra Research Institute, Chhatnag Road, Jhunsi,
%Allahabad 211 019, India}

\author{Junde Wu}
%\email{akpati@hri.res.in}

%\author{Uttam Singh}
%\email{uttamsingh@hri.res.in}

\affiliation{Department of Mathematics\\
Zhejiang University, Hangzhou 310027, PR~ China}

%\author{}
%\email{}

%\affiliation{}

\date{\today}

\begin{abstract}
We prove uncertainty relations that quantitatively express the impossibility of jointly sharp 
preparation of pre- and post-selected quantum states for measuring incompatible observables during 
the weak measurement. By defining a suitable operator whose average in the pre-selected quantum state 
gives the weak value,
we show that one can have new uncertainty relations for variances of two such operators corresponding to
two non-commuting observables. These generalize the recent stronger uncertainty relations that
give  non-trivial lower bounds for the sum of variances of two observables which 
fully capture the concept of incompatible observables.
Furthermore, we show that weak values for two non-commuting projection operators obey a
complementarity relation. Specifically, we show that for a pre-selected state if we measure a projector 
corresponding to an observable $A$ weakly followed by the strong measurement of another observable $B$ 
(for the post-selection) and, for the same pre-selected state we measure a projector corresponding 
to an observable $B$ weakly followed by the strong measurement of the observable $A$ (for the post-selection),
 then the product of these two weak values is always less than one. This shows that even though individually they are complex and can be large, their product
is always bounded. 
%This can have implication in measurement of wavefunction in complementary bases.
\end{abstract}

\maketitle

\section{ Introduction}
Quantum theory has many counter intuitive features such as the wave-particle duality, interference, 
entanglement and non-locality, and 
%that is what it makes
these features make the subject 
exciting even after ninety years of its initial formulation. To this weird list, the weak value adds another twist 
making quantum theory even more stranger than before.
The concept of weak value was introduced by Aharonov-Albert-Vaidman
\cite{aha,av} while investigating the properties of a quantum system in pre and post-selected ensembles.
 If the system is weakly coupled to an apparatus, then upon post-selection of the system state, 
 the apparatus variable is shifted by a weak value.
 The weak value can have strange properties. In particular, it can be a complex number, in general,
 and can take values outside the spectrum of the observable being measured. 
 This is in sharp contrast to the average value of an observable which is always bounded by the min- and max- of 
 its eigenvalue. This gives rise to 
 the notion of anomalous weak value for an observable. However, the precise sense in which the weak
 value is anomalous is defined only recently \cite{akp}.
 %In particular, if one measures an observable $O$ of the system weakly,
%with a preselected state $\ket{\psi_i}$ at time $t_i$ and postselected state
%$\ket{\psi_f}$ at time $t_f$, the value of the observable measured at time $t_i\leq t \leq t_f$ is
%given by the weak value of observable, which is $\braket{\psi_f}{O|\psi_i}/\braket{\psi_f}{\psi_i}$.
The concept of weak measurements has been generalized in various directions 
\cite{sw1,ADL,sw2,nori,shik}  
and have found numerous applications \cite{eric,sandu,cho,wise1,mir,js,ams1,ams2,pdav,jsl1,hof,jd}.

In quantum theory uncertainty relations play fundamental roles and have dominated the developments
of physics that ranges from foundations to quantum information, quantum communication and other
areas as well. This continues to be at center stage even after so many years of development of
quantum theory \cite{heis,ken,whz,rob,la,lah,ak,mo1,mo2,mo3,mo4,je,lar,bra,cbra,rfw,lm,rfw1,rfw2,rfw3,bus}. 
%(More to be added...) 
In one hand, there is the Robertson uncertainty relation \cite{rob} that is supposed to capture 
the preparation uncertainty about the 
quantum ensemble and the other one is based on the measurement-disturbance principle 
\cite{mo1,mo2,mo3,mo4,je,lar,bra,cbra,rfw,lm,rfw1,rfw2,rfw3,bus}.
The later principle tries to formalise the original thought experiment of Heisenberg quantitatively.
One should emphasize here that the impossibility of jointly sharp preparation of quantum ensemble and 
 the impossibility of jointly precise measurements of incompatible
observables are independent concepts.
Traditionally the uncertainty relations are expressed in terms of the product of variances of the measurement results of 
two non-commuting observables. Unfortunately, the Heisenberg-Robertson relation does not really capture 
the incompatible nature of two operators as the inequality can be trivial for some quantum sates. That is to say, 
that the product of variances can be null even when one of the two variances is
different from zero. Recently, two stronger uncertainty relations have been proved which go beyond the
Heisenberg-Robertson uncertainty relation \cite{lorenzo}. 
The new relations provide nontrivial bounds whenever the observables are incompatible
on the quantum state.

Here we ask can one have uncertainty relations that will impose fundamental 
limitations on the preparation of the pre- and post-selected (PPS) ensembles while measuring two non-commuting observables 
during the weak measurement? This question has been paid no or little attention in the context of weak measurements.
%Therefore, we explore if one can have the uncertainty relation for two non-commuting observables during the weak measurement.
In the usual strong measurement scenario, if we perform measurement of an observable then 
we can obtain an eigenvalue of the observable with some probability and the experimental
error bars are connected with the standard variance of the corresponding observable.
However, when we perform weak measurement of an observable, we do not get an eigenvalue rather a
complex number--the `weak value'. Therefore, to formulate the uncertainty relation during the weak measurement of two 
non-commuting observables, we have to define 
the variance of an operator such that its average over the pre-selected ( or post-selected) state gives us the weak value.
Further, the operator cannot be a Hermitian one, as it should give a complex number, in general.
Thus, such an operator has to be non-Hermitian and we have to define the variance of a non-Hermitian operator.
Indeed, by defining a suitable non-Hermitian operator, we will see that its average over the pre-selected
state gives the weak value and that will help to define the associated variance. Then, we prove two
new uncertainty relations for these non-Hermitian operators corresponding to
incompatible observables. They generalize the recent stronger uncertainty relations that
give  non-trivial lower bounds for all incompatible observables.
Further, we show that weak values for two non-commuting projectors obey a
complementarity relation. In particular, if we measure a projector corresponding to an observable $A$
weakly and  an observable $B$ strongly
(for the post-selection) and next time, for the same pre-selected state we reverse the order, 
i.e., we measure the projector 
corresponding to an observable $B$ weakly and an observable $A$ strongly (for the post-selection), 
then the product of these two weak values is always
less than one. This shows that even though they are complex and individually they can be large their product
is always bounded by one. The complementarity relation holds both in finite and infinite dimensional Hilbert 
spaces. Thus, the complementarity of weak values of two non-commuting projectors in a general feature of quantum system.
% The physical meaning of the complementarity relation is the following. 
The weak values corresponding to two projectors act like gentle windows for mutually exclusive features of a
quantum system. However, the complementarity relation suggests that both windows cannot be large at the same time.

The paper is organized as follows. In section II, we define the non-Hermitian operator whose average in the pre-selected
state gives the desired weak value for the observable of interest and then prove two new uncertainty relations for the non-Hermitian 
operators that represent weak values for two non-commuting operators. In section ~~III, we present the complementarity 
relations for weak values corresponding to two non-commuting projectors. Finally, we conclude in section IV.

%\section{ Weak operator and weak value}
\section{ Uncertainty relations during weak measurement}
We start with a system which is preselected in the quantum state $\ket{\psi_i} = \ket{\psi}$ and we are 
interested in measuring
%$A$ in state $\ket{\psi}$
an observable of the system weakly.
The weak measurement can be realized using the interaction between the system and the
measurement apparatus which is governed by the interaction Hamiltonian
\begin{align}
 H_{int}= g \delta(t-t_0) A \otimes M,
\end{align}
where $g$ is the strength of the interaction that is sharply peaked at $t=t_0$, $A$ is an observable of the system and $M$ is that of the apparatus.
This is the von Neumann model of measurement when the coupling strength is arbitrary. But if $g$ is small, then we can realize the weak
measurement \cite{aha,av}. The interaction Hamiltonian allows the system and apparatus to evolve as
\begin{align}
 \ket{\psi} \otimes \ket{\Phi}  \rightarrow e^{-\frac{i}{\hbar}g A \otimes M} \ket{\psi} \otimes \ket{\Phi}.
\end{align}
After the weak interaction, we postselect the system in the state $\ket{\phi}$ with the
postselection
probability given by $p= |\braket{\phi}{\psi}|^2 (1 + 2 g Im \langle A\rangle_w \langle M \rangle)$. This yields the desired weak value of $A$ as
given by
\begin{align}
\langle A\rangle_w = \frac{\braket{\phi}{A |\psi }}{\braket{\phi}{\psi} }.
\end{align}
The weak value can be inferred from the final state of the apparatus. After the post-selection of the system state, the 
final state of the apparatus (un-normalized) is given by
\begin{align}
 \ket{\Phi_f} = e^{-\frac{i}{\hbar}g  \langle A \rangle_w  M} \ket{\Phi} = \sum_m \phi_m  e^{-\frac{i}{\hbar}g  \langle A \rangle_w  m}
 \ket{m},
\end{align}
where $\ket{\Phi} = \sum_m \phi_m \ket{m}$ and the meter observable satisfies the eigenvalue equation $M\ket{m} = m\ket{m}$.

For a given pre- and post-selected ensemble, 
let us define the operator $A_w$ as
\begin{align}
A_w = \frac{\Pi_{\phi} A}{k},
\end{align}
where $\Pi_{\phi} = \ket{\phi} \bra{\phi}$ and $k = |\braket{\phi}{\psi}|^2 $. This has following properties: (i) it is 
non-Hermitian as $[\Pi_{\phi}, A] \not= 0$, (ii) $A_w \ket{\psi} = \langle A\rangle_w c \ket{\phi}$, where $c = \frac{1}{\braket{\psi}{\phi}}$,
and (iii) $A_w \ket{\phi} = \frac{\langle A\rangle}{k}  \ket{\phi}$, where $\langle A\rangle = \langle \phi| A|\phi \rangle$. 
%These properties suggest that
Thus, 
the weak value can be written as the average of the non-Hermitian operator $A_w$ in the state $\ket{\psi}$, i.e.,
$\langle A\rangle_w = \braket{\psi}{A_w |\psi }$.
%Note that the weak operator is non-Hermitian.
Moreover, the final post-selected state $\ket{\phi}$ is an eigenstate of the operator $A_w$ 
with the eigenvalue $\braket{\phi}{A| \phi}/k$.

Now, we prove that for two non-commuting operators $A$ and $B$,
their corresponding non-Hermitian operators $A_w$ and $B_w$ respect  non-trivial uncertainty relations for the
pre- and post-selected states $\ket{\psi}$ and $\ket{\phi}$.
The uncertainty relations presented here quantitatively express the impossibility of jointly
sharp preparation of pre- and post-selected quantum states for the weak measurement of incompatible 
observables. The first uncertainty relation that captures the limitation on the preparation of 
PPS ensemble is given by

\begin{align}
 \Delta A_w^2 + \Delta B_w^2 &\ge \pm \frac{i}{k} \bra{\phi}[A, B] \ket{\phi} \nonumber\\
& \pm i(\langle A\rangle_w^* \langle B\rangle_w - \langle A\rangle_w \langle B\rangle_w^*)\nonumber\\
&+ |\bra{\psi} ( A_w \pm i B_w ) \ket{\bar{\psi} } |^2
\end{align}
and the second uncertainty relation is given by

\begin{align}
 \Delta A_w^2 + \Delta B_w^2 &\ge
\frac{1}{2} |\bra{\psi} ( A_w + B_w ) \ket{\bar{\psi} } |^2.
\end{align}

To prove these two relations we need 
the concept of uncertainty for any general (non-Hermitian) operator $O$ in a quantum state $\ket{\psi}$ which can be defined as
\cite{ja,pati}
\begin{align}
\label{un}
 \Delta O^2 := \langle \psi|(O-\langle O\rangle) (O^\dagger-\langle O^\dagger\rangle)| \psi \rangle.
\end{align}
The above definition is motivated from the fact that if the operator is Hermitian, we should get back the usual expression for the
variance of the operator in the given state $\ket{\psi}$.
Since the operator $A_w$ is non-Hermitian, we can use (\ref{un}) to define the variance of the non-Hermitian operator $A_w$ in the state $\ket{\psi}$ as given by
\begin{align}
 \Delta A_w^2 := \langle \psi|(A_w-\langle A_w\rangle) (A_w^\dagger-\langle A_w^\dagger\rangle)| \psi \rangle,
\end{align}
where $\langle A_w\rangle = \bra{\psi}A_w\ket{\psi}$ and $\langle A_w^{\dagger} \rangle = \bra{\psi}A_w^{\dagger}\ket{\psi} = 
\langle A_w\rangle^*$ . Note that $\Delta A_w^2$ can also be written as
\begin{align}
 \Delta A_w^2 = \bra{\psi}A_w A_w^{\dagger} \ket{\psi} -\bra{\psi} A_w \ket{\psi} \bra{\psi} A_w^\dagger \ket{\psi}.
\end{align}
Similarly, for an observable $B$, we can define the operator $B_w = \frac{\Pi_{\phi}B}{k} $. 
Then, we have the uncertainty for $B_w$ in the pre-selected state $\ket{\psi}$ as
\begin{align}
 \Delta B_w^2 = \bra{\psi} B_w B^{\dagger}  \ket{\psi} -\bra{\psi} B_w \ket{\psi} \bra{\psi} B_w^{\dagger} \ket{\psi}.
\end{align}

Even though these operators $A_w$ and $B_w$ are non-Hermitian, their uncertainties in the state $\ket{\psi}$ is a positive and 
real number.
Also, we see that if the post-selected state $\ket{\phi}=\ket{\psi}$, then 
the uncertainty in the non-Hermitian operator reduces to
the uncertainty in the observable $A$. (One may note that the weak value becomes the expectation value if $\ket{\phi}=\ket{\psi}$.)

Now, define two unnormalized vectors $\ket{f} = (A_w^{\dagger} -\langle A_w^{\dagger}\rangle)\ket{\psi} = C^{\dagger}|\ket{\psi}$
and $\ket{g} = (B_w^{\dagger} -\langle B_w^{\dagger}\rangle)\ket{\psi} = D^{\dagger}|\ket{\psi}$, where $C= A_w - \langle A_w\rangle$ and
$D= B_w - \langle B_w\rangle$. With these definitions we have
$|| f||^2 = \Delta A_w^2$ and $||g ||^2 = \Delta B_w^2$. Next, 
consider the parallelogram law in the system Hilbert space which is given by
\begin{align}
2 \Delta A_w^2 + 2\Delta B_w^2 =
 ||( C^{\dagger} + \alpha D^{\dagger} ) \ket{\psi } ||^2 + ||( C^{\dagger} - \alpha D^{\dagger} ) \ket{\psi } ||^2
\end{align}
with $\alpha \in {\mathbf C} $ and $|\alpha|=1$.
By choosing $\alpha = \pm i$, the norm $||\ket{f} \pm i \ket{g} ||^2= ||( C^{\dagger} \pm i D^{\dagger}) \ket{\psi} ||^2$ can be expressed as
\begin{align}
 ||( C^{\dagger} \pm i D^{\dagger}) \ket{\psi} ||^2 =  \Delta A_w^2 + \Delta B_w^2 \nonumber\\
 \pm i \bra{\psi} (C D^{\dagger} - D C^{\dagger}) \ket{\psi}.
\end{align}
The last term can be simplified as
\begin{align}
  \pm i \bra{\psi} (C D^{\dagger} - D C^{\dagger}) \ket{\psi} = \pm \frac{i}{k} \bra{\phi}[A, B] \ket{\phi} \nonumber\\
  \pm i(\langle A\rangle_w^* \langle B\rangle_w - \langle A\rangle_w \langle B\rangle_w^*).
\end{align}
Now consider the quantity $|\bra{\psi} ( A_w \pm i B_w) \ket{\bar{\psi} } |^2 $, where $\ket{\bar{\psi} } $ is orthogonal to $\ket{\psi} $.
This can be expressed as
\begin{align}
& |\bra{\psi} ( A_w \pm i B_w ) \ket{\bar{\psi} } |^2 \nonumber\\
& = |\bra{\psi} ( A_w \pm i B_w ) -
( \langle A \rangle_w \pm i \langle B \rangle_w )  \ket{\bar{\psi} } |^2 \nonumber\\
& =  |\bra{\psi} ( C \pm i D ) \ket{\bar{\psi} } |^2.
\end{align}
Using the Cauchy-Schwarz inequality, we have
\begin{align}
 |\bra{\psi} ( C \pm i D ) \ket{\bar{\psi} } |^2 \le || ( C^{\dagger} \mp i D^{\dagger} ) \ket{\psi } ||^2.
\end{align}
Therefore, for two operators $A_w$ and $B_w$ we have the uncertainty relation as given by
\begin{align}
 \Delta A_w^2 + \Delta B_w^2 &\ge \pm \frac{i}{k} \bra{\phi}[A, B] \ket{\phi} \nonumber\\
& \pm i(\langle A\rangle_w^* \langle B\rangle_w - \langle A\rangle_w \langle B\rangle_w^*)\nonumber\\
&+ |\bra{\psi} ( A_w \pm i B_w ) \ket{\bar{\psi} } |^2.
\end{align}
This is the first uncertainty relation that we wanted to prove which quantitatively expresses the impossibility of joint sharp preparation 
of PPS ensemble for two non-commuting observables during the weak measurement.
Note that the term containing the commutator between $A$ and $B$ appears as averaged in the final post-selected state.
This is consistent with the fact that in order to
obtain general weak values for $A$ and $B$, the post-selected state should not be an eigenstate of these observables.
%Hence this fact is taken care by this term. 
The second term arises due to the non-Hermitian nature of the operators $A_w$ and $B_w$. For 
Hermitian operators, this term is precisely zero. 
We should chose the sign of the first and second term so that they are real and positive. This is also a generalized stronger version of the uncertainty relation \cite{lorenzo} 
in the sense that if $\ket{\psi}$ is an eigenstate of one of the observable (either $A$ or $B$) still we have a
non-trivial lower bound for the uncertainty of the other operator.

 For canonical conjugate pair of observables, such as the position and the momentum operator of a particle the above 
 uncertainty relation takes the form 
 \begin{align}
 \Delta X_w^2 + \Delta P_w^2 &\ge \frac{\hbar}{k} \pm 2 Im (\langle X\rangle_w \langle P\rangle_w^*) \nonumber\\
&+ \frac{2}{k} |\bra{\phi} ( a^{\dagger} ) \ket{\bar{\psi} } |^2,
\end{align}
where $\langle X\rangle_w,  \langle P\rangle_w$ are the weak values of the position and momentum observables, respectively and
$a^{\dagger}$ is the creation operator.

Indeed, if we take the post-selection state $\ket{\phi} = \ket{\psi}$, then we have the stronger 
uncertainty relation for two Hermitian operators
$A$ and $B$  as given by \cite{lorenzo}
\begin{align}
 \Delta A^2 + \Delta B^2 &\ge \pm i \bra{\psi}[A, B] \ket{\psi} \nonumber\\
&+ |\bra{\psi} ( A \pm i B ) \ket{\bar{\psi} } |^2.
\end{align}
%To see how strong In contrast to the Heisenberg-Robertson uncertainty relation
The stronger uncertainty relation for two canonical conjugate pair of observables, such as the position and the 
momentum operator of a particle is given by
 \begin{align}
 \label{MP}
 \Delta X^2 + \Delta P^2 \ge \hbar + 2 |\bra{\phi} ( a^{\dagger} ) \ket{\bar{\psi} } |^2,
\end{align}
Note that the Heisenberg-Robertson uncertainty relation for the position and momentum implies that we have 
$\Delta X^2 + \Delta P^2 \ge \hbar$. Hence, the new uncertainty relation \cite{lorenzo} 
goes beyond the usual Heisenberg-Robertson uncertainty relation. This shows that quantum 
uncertainties given in (\ref{MP}) contains more intrinsic randomness of the outcomes of quantum tests performed on identically prepared 
quantum states than what is allowed by the Heisenberg uncertainty relation.

Next, we prove another uncertainty relation for two operators $A_w$ and $B_w$ which can be stated as
%Therefore, for two weak operators we have the uncertainty relation
\begin{align}
 \Delta A_w^2 + \Delta B_w^2 &\ge
\frac{1}{2} |\bra{\psi} ( A_w + B_w ) \ket{\bar{\psi} } |^2.
\end{align}

%To prove this let $\ket{f} = (A_w^{\dagger} -\langle A_w^{\dagger}\rangle)\ket{\psi} = C^{\dagger}|\ket{\psi}$
%and $\ket{g} = (B_w^{\dagger} -\langle B_w^{\dagger}\rangle)\ket{\psi} = D^{\dagger}|\ket{\psi}$ are two unnormalized
%vectors. 

To prove this consider the parallelogram law in the Hilbert space with $\alpha =1$. This leads to 
\begin{align}
2 \Delta A_w^2 + 2\Delta B_w^2 =
 ||( C^{\dagger} + D^{\dagger} ) \ket{\psi } ||^2 + ||( C^{\dagger} - D^{\dagger} ) \ket{\psi } ||^2.
\end{align}
One can check that
\begin{align}
||( C^{\dagger} + D^{\dagger} ) \ket{\psi } ||^2 = \Delta^2(A_w+B_w),
\end{align}
 where $\Delta^2(A_w+B_w)= \bra{\psi} (A_w + B_w)( A_w^{\dagger} + B_w^{\dagger}) \ket{\psi} -
 \bra{\psi} (A_w + B_w) \ket{\psi} \bra{\psi}( A_w^{\dagger} + B_w^{\dagger}) \ket{\psi}$ is the uncertainty in the sum of
 two weak operators. From the above, we have $2 \Delta A_w^2 + 2\Delta B_w^2 \ge \Delta^2(A_w+B_w)$.

  Given any non-Hermitian operator $O$,
 we can show that its action on a state can be expressed as
\begin{align}
O^{\dagger}  \ket{\psi }  = \langle O^{\dagger}  \rangle \ket{\psi} +  \Delta O \ket{\bar{\psi_O}}.
\end{align}
where $\langle O^{\dagger}  \rangle =  \bra{\psi} O^{\dagger}  \ket{\psi }$,
$\Delta O$ is as defined in (\ref{un}) and $\braket{\psi}{\bar{\psi_O}} =0$. 
The above formulae is a generalization of the Vaidman formula \cite{vaid} for any non-Hermitian operator.
Hence, we can define the action of the  operator $( A_w^{\dagger} + B_w^{\dagger})$ on any state as
\begin{align}
( A_w^{\dagger} + B_w^{\dagger})  \ket{\psi }  = (\langle A \rangle_w^* + \langle B \rangle_w^*)\ket{\psi } 
+  \Delta ( A_w + B_w) \ket{\bar{\psi}},
\end{align}
where $\ket{\bar{\psi}} \in {\bar {\cal H}}$  is orthogonal to the state $\ket{\psi}$.
Using the above formula we can express $\Delta^2(A_w+B_w) = |\bra{\bar{\psi}}( A_w^{\dagger} + B_w^{\dagger}) \ket{\psi}|^2 =
|\bra{\psi}( A_w+ B_w) \ket{\bar{\psi}}|^2$. Hence the proof.

Again, we can check that if we take the post-selection state $\ket{\phi} = \ket{\psi}$, then we have the stronger
uncertainty relation for two Hermitian operators as given by \cite{lorenzo} 
\begin{align}
 \Delta A^2 + \Delta B^2 &\ge
\frac{1}{2} |\bra{\psi} ( A + B ) \ket{\bar{\psi} } |^2
\end{align}

The uncertainty relations presented in this section captures the impossibility of joint sharp preparation of PPS ensemble 
for two incompatible observables during the weak measurement.

\section{Complementarity of weak values}
In quantum theory complementarity imposes limitation on our ability to display two mutually exclusive
properties of a quantum system at the same time. The principle of complementarity for the 
wave-particle duality as formulated by Bohr 
is a famous example of this \cite{whz}. In the context of weak measurement, it is possible that one can probe two
complementary aspects of a quantum system at some price (for e.g. introducing more noise) as the apparatus 
interacts with the system weakly allowing
 a gentle observation without disturbing the system too much. However,
we will show that if one performs weak measurements
of two non-commuting projectors in the reverse order between two strong measurements, then 
there is a complementarity relation that holds for weak values in the context of weak measurement.

Consider $A$ and $B$ as two non-commuting operators with $A= \sum_a a \Pi_a$ and $B= \sum_b b \Pi_b$ being their
spectral decompositions, respectively. Let the system is pre-selected in the state $\ket{\psi}$ and we perform the weak measurement of
the projector $\Pi_a$ with the post-selection in the state $\ket{\phi} =\ket{b}$, i.e., measure the observable $B$ strongly.
In this case, the weak value is given by 
%\begin{align}
$\langle \Pi_a \rangle_w^{(b)} = \frac{\bra{b} \Pi_a \ket{\psi}}{\bra{b} \ket{\psi}}$.
%\end{align}
 Instead, suppose
we perform the weak measurement of the projector $\Pi_b$ with the  post-selection in the state $\ket{\phi} =\ket{a}$, 
i.e., measure the observable $A$ strongly.
In this case, the weak value is given by 
$\langle \Pi_b \rangle_w^{(a)} = \frac{\bra{a} \Pi_b \ket{\psi}}{\bra{a} \psi \rangle}$.
From these definition, we see that the weak value actually connects wavefunctions directly in complementary bases. For example, we have
\begin{align}
\langle \Pi_a \rangle_w^{(b)} \psi(b)  &= \bra{b} a \rangle \psi(a), \nonumber\\
 \langle \Pi_b \rangle_w^{(a)} \psi(a) & = \bra{a} b \rangle \psi(b),
\end{align}
where $\psi(a)$ and $\psi(b)$ are the wavefunctions in the eigenbasis representation of $A$ and $B$, respectively.
The interesting point to note here is $\psi(a)$ and $\psi(b)$ are directly related without the unitary transformation. This simple
 observation illustrates the importance of weak value, that how they connect two complementary aspects directly.
 This is as if the weak value $\langle \Pi_a \rangle_w^{(b)}$ acts as a window to pop into $ \psi(b)$ and reveal 
 $\psi(a)$ and similarly, the weak value $\langle \Pi_b \rangle_w^{(a)}$ acts as a window to pop into $ \psi(a)$ and reveal
 $\psi(b)$.

 Next, we ask can these two weak values be arbitrarily large at the same time? Strangely, not.
 First, note that the weak values for the projectors $\Pi_a$ and $\Pi_b$ can be expressed as
 the sum of the average of the projectors in the state $\ket{\psi}$ plus an anomalous part \cite{akp}
 \begin{align}
\label{vaid}
\langle \Pi_a\rangle_w^{(b)} &= \langle \Pi_a \rangle  +  \Delta \Pi_a \frac{ \braket{b}{ {\bar{\psi_a}}} }{ \braket{b}{\psi} }, \nonumber\\
\langle \Pi_b\rangle_w^{(a)} &= \langle \Pi_b \rangle  +  \Delta \Pi_b \frac{ \braket{a}{ {\bar{\psi_b}}} }{ \braket{a}{\psi} },
\end{align}
 where $\langle \Pi_a \rangle =  \bra{\psi} \Pi_a  \ket{\psi }$,
$\Delta \Pi_a$ is the uncertainty of the projector in the state $\ket{\psi}$, i.e., 
$\Delta \Pi_a^2 = \langle \psi|(\Pi_a-\langle \Pi_a\rangle)^2 |\psi \rangle = |\psi(a)|^2 (1- |\psi(a)|^2)$
and $\ket{\bar{\psi_a}} \in {\bar {\cal H}}$ is a state orthogonal to $\ket{\psi}$. Similarly meanings 
hold for the other projector $\Pi_b$.
This shows that the weak values of these projectors can be large and remain outside the range of these observables.
However, both these weak values cannot be large at the same time for the same pre-selected state.
Indeed, using the Cauchy-Schwarz inequality, we see that the product of these two weak values satisfies
\begin{align}
\langle \Pi_a \rangle_w^{(b)}   \langle \Pi_b \rangle_w^{(a)} = |\bra{a} b \rangle|^2  \le 1.
\end{align}

This shows that even though individually each of these weak values can be arbitrarily large, their product is independent of the
pre-selected state and is bounded by one.
%This arises due to the non-commuting nature of the operators involved.
This is a new kind of complementarity for quantum weak values of two non-commuting observables. This feature is akin to
quantum world. One cannot imagine such a complementarity relation in the classical world.
%One may say that classically, this cannot happen as they can be individually large and their products can be also

 Similar complementarity also holds for two non-commuting observables such as the position and the momentum projectors
 in the infinite dimensional Hilbert space.  First, note that in infinite dimensional Hilbert space, the projection operator
 for the position observable is not $\ket{x}\bra{x}$. If we define the projection operator $\Pi_x = \ket{x}\bra{x}$, then
  it is ill defined and it will not satisfy the condition 
$\Pi_x^2 = \Pi_x$ as the position eigenstates satisfy $\braket{x}{x'} =\delta(x -x')$. The correct projection operator for continuous 	
variables is defined as 	
\begin{align}
\label{pro}
\Pi_{\Delta x} = \int_{x_0 -\frac{\Delta x}{2}}^{x_0 + \frac{\Delta x}{2}} ~dx'  \ket{x'}\bra{x'}.
\end{align}	
The reason for this definition is that we cannot project 	
an arbitrary state which is represented in terms of continuous basis state onto a point to get the exact eigenvalue. 	
There will be always a spread within an interval. We can 	
only project a state around $x_0$ to a selectivity $\Delta x$ of the 	
measuring apparatus. It is not possible to design a device 	
to make a perfectly selective measurement of a continuous variable.
The interval in a continuous spectrum cannot be narrowed 	
down, because it will always contains an infinite number of eigenvalues.
Thus, if we have a wave packet the effect of projection is to truncate it around $x_0$ within 	
an interval $\Delta x$. The operator defined in (\ref{pro}) satisfies $\Pi_{\Delta x}^2 = \Pi_{\Delta x}$ and
$\Pi_{\Delta x}\ket{x} = \ket{x}$, thus indicating that it is indeed a valid projection operator.
Similarly, we can define a projection operator for the momentum observable as
\begin{align}
\Pi_{\Delta p} = \int_{p_0 -\frac{\Delta p}{2}}^{p_0 + \frac{\Delta p}{2}} ~dp'  \ket{p'}\bra{p'}
\end{align}	
and this indeed satisfies $\Pi_{\Delta p}^2 = \Pi_{\Delta p}$ and
$\Pi_{\Delta p}\ket{p} = \ket{p}$.

Let us consider $\Pi_{\Delta x}$ and $\Pi_{\Delta p} $ as two non-commuting projectors, corresponding to 
position and momentum of a particle, respectively.
 Let the system of interest is pre-selected state in the state $\ket{\psi}$ and we perform the weak measurement of
the projector $\Pi_{\Delta x}$ with post-selection in the momentum eigenstate $\ket{\phi} =\ket{p}$ 
(i.e., we measure the observable $p$ strongly). In this case, the weak value for the projection operator of position is given by
\begin{align}
\langle \Pi_{\Delta x} \rangle_w &=  \frac{\bra{p} \Pi_{\Delta x} \ket{\psi}}{\braket{p}{\psi}} \nonumber\\
&=  \frac{1}{\sqrt{2\pi \hbar}} \int_{x_0 -\frac{\Delta x}{2}}^{x_0 + \frac{\Delta x}{2}} ~dx e^{-i px/\hbar}
\frac{\psi(x)}{\psi(p)}.
\end{align}
As a side remark, one  can see that if one post-selects in a momentum state $\ket{p=0}$, then the above weak value gives
$\langle \Pi_{\Delta x} \rangle_w = C \int_{x_0 -\frac{\Delta x}{2}}^{x_0 + \frac{\Delta x}{2}} ~dx
\psi(x)$, where $C = \frac{1}{ \sqrt{2\pi \hbar} \psi(p=0)}$. Note that this result is different than 
Lundeen {\it et al} \cite{jsl1} where they
have taken the projection operator for the position observable simply as $\ket{x}\bra{x}$. Even though this is well defined for 
the discrete basis, it is
not well defined for the continuous basis. The weak value of the position projection operator with the pre-selected state 
$\ket{\psi}$ and post-selected state $\ket{p=0}$ actually gives an integrated wavefunction around
$x_0$ with a selectivity $\Delta x$. However, we do not go into detail discussions about the measurement of 
quantum wavefunction using the weak measurement as that is not our main aim here.

To show the complementarity relation,  instead, suppose
we perform the weak measurement of the projector $\Pi_{\Delta p}$ with the post-selection being performed 
in the position eigenstate $\ket{\phi} =\ket{x}$ (i.e.,  we measure the observable $x$ strongly).
In this case, the weak value of the  projector $\Pi_{\Delta p}$ is given by 
\begin{align}
\langle \Pi_{\Delta p} \rangle_w &= \frac{\bra{x} \Pi_{\Delta p} \ket{\psi}}{\braket{x}{\psi}} \nonumber\\
& = \frac{1}{\sqrt{2\pi \hbar}} \int_{p_0 -\frac{\Delta p}{2}}^{p_0 + \frac{\Delta p}{2}} ~dp e^{i px/\hbar}
\frac{\psi(p)}{\psi(x)}.
\end{align}
Again, we can check that the product of these two weak values satisfy the condition 
\begin{align}
\langle \Pi_{\Delta x} \rangle_w  \langle \Pi_{\Delta p} \rangle_w = 1.
\end{align}

Thus, the complementarity of weak values of two non-commuting projectors in a general feature of quantum system 
that holds both in the finite and infinite dimensions. The physical meaning of the complementarity relation is the 
following. The weak values corresponding to two projectors act like gentle windows for mutually exclusive features of a 
quantum system. However, the windows cannot be large for both complementary world at the same time.

\section{Conclusions}
In the standard interpretation of quantum theory, the  Heisenberg- Robertson uncertainty relation tells 
about the impossibility of preparing an ensemble of identically prepared quantum states such that the variances 
of two incompatible observables will be arbitrarily large. Rather, their product is bounded by the modulus of their commutator 
in a given quantum state.  In the context of weak measurements of two non-commuting observables, till date there is no  
uncertainty relation that imposes limitations on the joint sharp preparation of pre- and post-selected ensemble. 
In this paper, we have proved two new uncertainty 
relations which precisely capture the impossibility of joint sharp preparation of the pre- and post-selected ensemble 
while measuring the weak value of two non-commuting operators.
Moreover, we have argued that if we formulate the uncertainty relation using the Robsertson type of uncertainty relation, 
then we do not get a non-trivial bound. The new uncertainty relations are based on the sum of the variances of two non-Hermitian operators
whose average give the weak values for two incompatible observables and whose lower bound 
is always non-trivial. We have also proved a complementarity relation for two non-commuting 
projectors for a given pre-selected quantum state. This shows that even though, individual weak values for the projectors can be large, the product of the weak values of these
projectors for any pre-selected state 
%and post-selected in two different final states corresponding to two observables 
is always bounded by one. This reveals a strange feature associated with weak values of incompatible observables during weak measurements.

%\subsubsection*{Acknowledgement}
\vskip 1cm

\noindent
{\bf Acknowledgement:}
This work is supported by National Natural Science
Foundation of China (11171301 and 10771191) and the Doctoral Programs
Foundation of Ministry of Education of China (J20130061), and it is also supported
by the Special Project of University of Ministry of Education of China and the
Project of K. P. Chair Professor of Zhejiang University of China. We thank S. Wu for
useful remarks.

\end{document}